\documentclass[journal]{vgtc}                     

\onlineid{1249}

\vgtccategory{Research}

\vgtcpapertype{Representation \& Interaction}

\title{Unveiling the Visual Rhetoric of Persuasive Cartography: \\ A Case Study of the Design of Octopus Maps}

\author{Daocheng Lin, Yifan Wang, Yutong Yang and Xingyu Lan
}

\authorfooter{
  \item
  	Daocheng Lin is with Iowa State University.
  	E-mail: daocheng@iastate.edu.
  \item
  	Yifan Wang is with Fudan University.
  	E-mail: wangyifanlea@gmail.com.

  \item Yutong Yang is with Shanghai Jiao Tong University.
  	E-mail: flora20@sjtu.edu.cn.

  \item Xingyu Lan is with Fudan University.
  E-mail: xingyulan96@gmail.com. Xingyu Lan is the corresponding author.

}

\abstract{
When designed deliberately, data visualizations can become powerful persuasive tools, influencing viewers' opinions, values, and actions. While researchers have begun studying this issue (e.g., to evaluate the effects of persuasive visualization), we argue that a fundamental mechanism of persuasion resides in rhetorical construction, a perspective inadequately addressed in current visualization research. To fill this gap, we present a focused analysis of octopus maps, a visual genre that has maintained persuasive power across centuries and achieved significant social impact. Employing rhetorical schema theory, we collected and analyzed 90 octopus maps spanning from the 19th century to contemporary times. We closely examined how octopus maps implement their persuasive intents and constructed a design space that reveals how visual metaphors are strategically constructed and what common rhetorical strategies are applied to components such as maps, octopus imagery, and text. Through the above analysis, we also uncover a set of interesting findings. For instance, contrary to the common perception that octopus maps are primarily a historical phenomenon, our research shows that they remain a lively design convention in today’s digital age. Additionally, while most octopus maps stem from Western discourse that views the octopus as an evil symbol, some designs offer alternative interpretations, highlighting the dynamic nature of rhetoric across different sociocultural settings. Lastly, drawing from the lessons provided by octopus maps, we discuss the associated ethical concerns of persuasive visualization.

}


\teaser{
  \centering
  \includegraphics[width=\linewidth, alt={Examples of octopus maps.}]{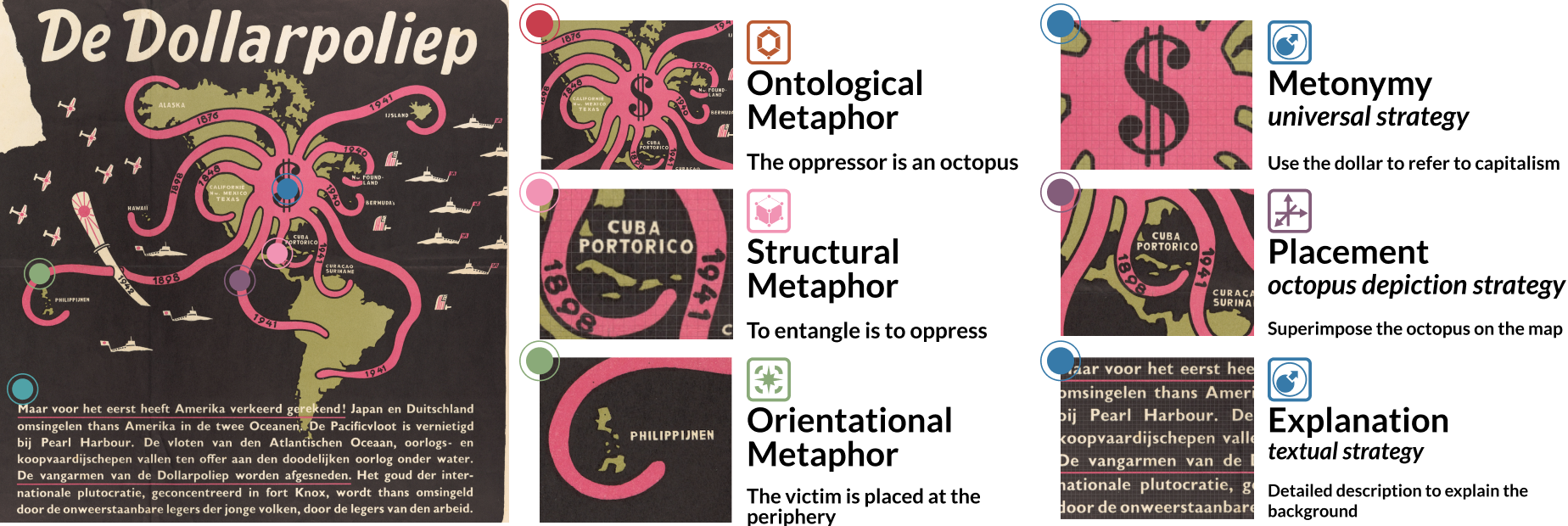}
  \caption{An octopus map satirizing the US from our corpus and part of the analysis codes we generated when performing visual rhetoric analysis on it. We introduced our coding framework in \cref{sec:framework} and the complete version of the derived design space in \cref{sec:space}.}
  \label{fig:teaser}
}

\graphicspath{{figs/}{figures/}{pictures/}{images/}{./}} 

\usepackage{tabu}                      
\usepackage{booktabs}                  
\usepackage{lipsum}                    
\usepackage{mwe}                       

\usepackage{mathptmx}                  

\newcommand{\etal}{et~al.~} 
\newcommand{\ie}{i.e.,~}
\newcommand{\eg}{e.g.,~}
\newcommand{\ncorpus}{90 }

\usepackage{graphicx}
\usepackage{wrapfig}
\usepackage[T1]{fontenc}
\usepackage{amsmath}  
\usepackage{multirow}
\usepackage{tabularx}
\newcommand{\x}[1]{{\leavevmode\color{black}{#1}}}

\usepackage[svgnames,x11names,table]{xcolor}
\usepackage{soul}
\usepackage{adjustbox}
\usepackage{lipsum}
\usepackage{url}

\begin{document}

\firstsection{Introduction}
\maketitle

Data visualization can be persuasive~\cite{pandey_persuasive_2014}. When designed with certain intents, visualizations possess the power to swing emotions, attitudes, and actions. For example, visualizations have a long history of being used by politicians and businessmen to convey information that is advantageous to them, through methods such as truncating axes and distorting data encodings~\cite{pandey2015deceptive,lan_i_2024}. 
During the COVID-19 pandemic, Lee~\etal~\cite{lee2021viral} found that both supporters and opponents of mask-wearing posted data visualizations on social media that supported their stances, in order to defend themselves or attack the opposing camp. Lisnic~\etal~\cite{lisnic2023misleading} found that COVID skeptics used various methods to lie with visualizations and spread misinformation. In addition, a series of user studies~\cite{garreton2024attitudinal,muehlenhaus2013design,heyer2020pushing,kong2018frames} have found that specific factors in visualization designs (\eg colors, titles, illustrations) can influence users' take-away messages and even lead to changes of attitudes.

However, although previous research has laid a solid foundation for our work, an important but under-explored perspective of studying persuasive visualization is visual rhetoric, which examines ``how images act rhetorically upon viewers''~\cite{hill2012defining}. Since Aristotle defined rhetoric as ``the ability to see the available means of persuasion in any given case'', rhetoric has been an integral part of persuasion. As Bourdieu noted~\cite{bourdieu1991language}, rhetoric functions to ``endows a group with a will, a
plan, a hope or, quite simply, a future''. This also applies to visual media such as data visualization. However, in the visualization community, relevant research is still limited. As one case, Hullman~\etal~\cite{hullman_visualization_2011} found various rhetorical frames in data stories. 
Campbell and Offenhuber~\cite{campbell2019feeling} studied the effect of proximity techniques (\ie making the data more relatable to the viewer) based on a rhetorical theory, which suggests that proximity in time and place can help enhance emotional appeal.
To further the research in this relatively under-explored area, this work aims to conduct an in-depth analysis on the design of persuasive visualizations to characterize their rhetorical techniques and strategies.

Specifically, we chose to focus on octopus maps as our research cases. Octopus maps are a visualization form highly popular in history~\cite{puerta2025many}. The common form is to depict a specific entity (\eg country) as an octopus, with its tentacles threatening others. We believe it is necessary to study them because: (i) According to previous literature, maps are a typical medium for embedding ideology and power~\cite{barton_ideology_1993,monmonier_how_2018,harley2008maps}, and have been extensively used in political propaganda, wars, social movements, and other persuasive activities. (ii) Octopus maps are a typical design genre within persuasive maps~\cite{tyner_persuasive_1982}. They were repeatedly created and published, and had a wide social impact during events such as the two World Wars. This provides us with a natural and vivid example to understand how persuasive visualization is used in real social events and how visualization design can trigger a huge social impact. (iii) Octopus maps contain rich design elements: geographical data visualizations form the base map, while more figurative elements, such as the octopus and other embellishments, add additional layers of meaning. These elements offer a rich array of materials for analyzing visual rhetoric. Although a recent work by Puerta \etal~\cite{puerta2025many} also turned their attention to octopus maps and inspiringly proposed six high-level features of octopus design (\eg centrality, tentacularity, grabbiness), we believe a more low-level visual rhetoric analysis is still needed to systematically understand the design of such maps.

Specifically, we constructed a corpus containing \ncorpus octopus maps from various open sources (\eg map libraries, forums, social media). Then, we employed rhetorical schema theory to code the images and derived a design space for characterizing their visual rhetoric, which includes two main dimensions: conceptual metaphors and rhetorical strategies. Based on this analysis, we reflected on the dynamics of visual rhetoric across cultures and times, and discussed the potential ethical issues associated with persuasive visualizations.

In a word, the contributions of this research include:
\begin{itemize}
    \item We curated a corpus of octopus maps from various online sources. Based on the analysis of this corpus, we derived a design space that systematically illustrates the visual rhetoric of octopus maps.
    \item We referred to the theories from rhetoric studies and provided additional theoretical perspectives for understanding the design construction of persuasive visualizations.
    \item We proposed observations and reflections based on the analysis of octopus maps, offering insights for researchers in the fields of data storytelling and visual communication, as well as for practitioners such as data journalists, visual designers, government communicators, and educators.
\end{itemize}

*This work may contain images that depict certain countries or regions in a highly political and ideological manner. However, these images are included solely for analytical purposes and do not reflect any political stance of the authors.


\section{Background and Related Work}




\subsection{Persuasive Visualization}

Visualization can function as a powerful communicative device and be used to persuade users~\cite{pandey_persuasive_2014,lee-robbins_affective_2023}. At the application level, the field of data storytelling has paid particular attention to persuasive visualization. For example, some data stories have been found to have the intent of advocating opinions or values~\cite{lan2023affective}, or to be involved in political agendas~\cite{lee-robbins_affective_2023}. Thus, researchers have inquired about the ``rhetorical power of visualizations'' in data stories~\cite{garreton2024attitudinal}. 
Empirically, Hullman and Diakopoulos~\cite{hullman_visualization_2011} examined 51 data stories and found various intentional frames and rhetoric applied to visualizations. 
Pandey~\etal~\cite{pandey_persuasive_2014} carried out an experiment to examine how visualization in data stories influences attitudinal change, suggesting that especially when participants do not possess a strong initial attitude, charts can be highly persuasive.
A series of studies have examined how titles~\cite{kong2018frames}, narrative structures~\cite{yang2021design}, anthropomorphism~\cite{errey2024nudging,morais2020showing}, and visual elements such as animation~\cite{shi2021communicating} and emphasis cues~\cite{kong2019understanding} can be strategically designed to communicate certain messages.
At the same time, groups such as data journalists have long paid attention to issues of deception and misrepresentation in visualizations, actively reflecting on how media distort and manipulate visualizations to persuade audiences~\cite{huff2023lie,cairo2019charts}.

Among various visualization types, the persuasive power of maps has been repeatedly discovered and verified. As early as last century, the geographical and rhetorical communities had a concentrated discussion on the issue, finding that maps have been extensively used for purposes such as propaganda. For example, Tyner~\cite{tyner_persuasive_1982} coined the term ``persuasive cartography'' and discussed the various ways in which maps can be manipulated (\eg through projections, colors, and text) to serve persuasive purposes. Monmonier~\cite{monmonier_how_2018} reviewed abundant cases where cartographers introduced distortions or even non-existent objects on maps to serve political, commercial, and other vested interests. 
Harley~\cite{harley2008maps} discussed the relationship between maps and power.
Barton~\cite{barton_ideology_1993} investigated the rules of inclusion and exclusion in persuasive maps and how they are related to ideology. 
In recent years, rhetorical designs have still been constantly found in the creation of various types of maps and the storytelling of geographic data~\cite{muehlenhaus2013design,roth2021cartographic}.

\subsection{Octopus Maps}

Octopus maps are a representative and popular design genre within persuasive maps~\cite{puerta2025many,mode_not_2017,baynton-williams_curious_2015,tyner_persuasive_1982}. 
In Western culture, the octopus often symbolizes a powerful and pervasive entity, frequently associated with negative connotations such as greed or dominance~\cite{stangl_geographic_2016}. This symbolism may be traced back to Western legends of sea monsters and the fear of maritime civilizations towards oceanic creatures~\cite{van2013sea,nigg2014sea}. Before being systematically utilized in propaganda maps, numerous artworks concerning octopuses had already existed, and works like Victor Hugo’s popular novel \textit{Toilers of the Sea} further disseminated the imagery of octopus as ``glue filled with hatred'', ``the pneumatic machine attacking you'' to the public~\cite{hugo1899toilers}. 
Octopus maps inherit this historical lineage. However, unlike general pictures and illustrations, octopus maps creatively integrate the octopus imagery with real geographic data. This has enabled them to go beyond mere artistic creation and reflect more realistic issues, such as wars and confrontations between nations. 

However, despite the significant influence of octopus maps, research on them remains relatively limited. Previous studies have primarily taken place in the humanities (\eg historical geography and rhetoric studies), with methods generally based on discussions of individual cases~\cite{mode_not_2017,baynton-williams_curious_2015}. Recently, Puerta \etal~\cite{puerta2025many} conducted a study based on their personal collection of octopus maps, identifying six general features (\eg centrality, tentacularity) of octopus imagery. They then conducted a user experiment to examine whether the implicit existence of these features will ``inspire similar conspiratorial or adversarial responses''. In this work, aligning with their call for ``a deeper research of visual rhetoric on the octopus map'', we adopt the perspective of rhetorical schema theory to conduct a more in-depth analysis of the rhetorical techniques in octopus maps. By employing the iconographic tracking method, we build a more formal dataset and offer a granular and systematic design analysis, examining not only octopus features but also map features and their interrelationships.

\begin{figure*}[t!]
    \centering
    \includegraphics[width=1\linewidth]{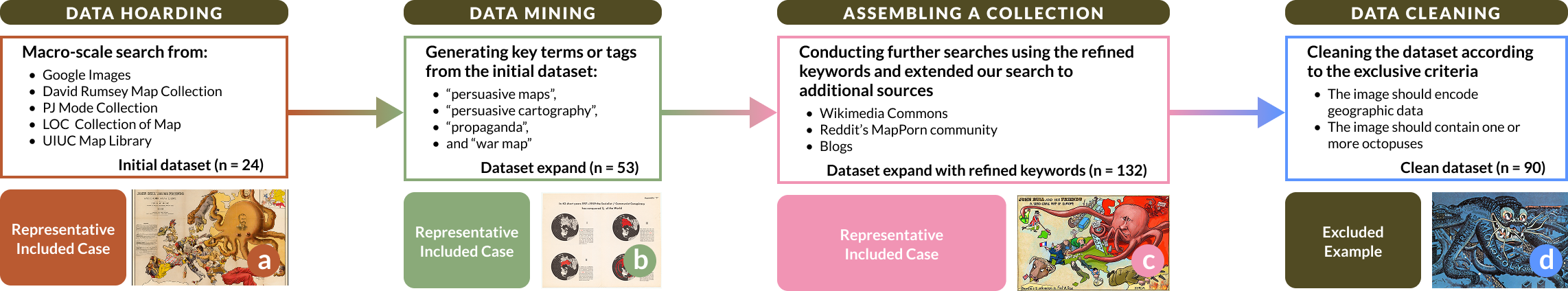}
    \caption{The pipeline of iconographic tracking.}
    \label{fig:criteriaOct}
    \vspace{-2em}
\end{figure*}

\subsection{Embellishments and Metaphors in Data Visualization}

The embellishments in visualizations (\ie pictorial elements such as icons and illustrations) have been a highly debated issue in the visualization community. For example, Tufte raised the famous concern of ``chartjunk''~\cite{few2011chartjunk}, while some researchers, through conducting user studies, have found that embellishments help enhance user engagement and memory~\cite{borgo_empirical_2012,bateman2010useful}. In recent years, the wide application of visual forms such as infographics has led some researchers to explore the design space of embellishments, such as the integration of icons and data marks in pictograms, and tools have been developed to facilitate expressive and stylish visualization design~\cite{xiao_let_2024, shi_supporting_2023}. 



Visual metaphor is a special embellishment technique. On the one hand, researchers have found that the incorporation of visual metaphors can aid data comprehension. For example, Huron~\etal~\cite{huron2013visual} leveraged the metaphor of sedimentation and proposed a novel visualization method to represent data streams.
Xie~\etal~\cite{xie_semantic-based_2019} utilized the galaxy visual metaphor to convey complex layers of information in large image collections. 
Meanwhile, visual metaphors are also communicative devices frequently found in forms such as data stories and artistic visualizations~\cite{lan2023affective, preim_survey_2024, kostelnickHumanizingVisualDesign2019,kimball_mountains_2016}. For instance, Lan~\etal~\cite{lan2022negative} discussed the well-known \textit{Iraq's bloody toll} design, which inverts a bar chart and colors it red to create a blood-like effect, noting that such metaphorical designs may be the deliberate choice of designers to attract attention or evoke emotional responses. However, they also found that users sometimes fail to decode visual metaphors, leading to confusion or misunderstanding. 
Although the role of embellishment and metaphor is still under discussion, overall, a gradual consensus is forming that in certain contexts (such as storytelling and persuasion), they are common and functional elements. Therefore, we believe that the octopus maps studied in this work provide valuable cases that enable us to understand how, at least in persuasive contexts, embellishments can be designed and assembled, and why they can have such a significant social impact.

\section{Curating A Corpus of Octopus Maps}
\label{sec:corpus}

Although many researchers have concluded that the octopus map is a highly representative and influential type of persuasive map in history~\cite{mode_not_2017, baynton-williams_curious_2015,tyner_persuasive_1982}, prior studies have generally focused on individual cases. To provide a more comprehensive overview of the design methods of octopus maps, in this section, we aim to construct a more systematic corpus of octopus maps from various available online sources.

\subsection{Methodology}
\label{ssec:corpus_method}


An exhaustive corpus of octopus maps is nearly impossible to collect since the ubiquitous propaganda can be ``rapidly discarded and destroyed'' in history~\cite{puerta2025many}. Moreover, since early octopus maps were mostly printed in concrete forms such as lithography or woodcut, the original versions are almost inaccessible, and only those that have been digitized are publicly available. Rhetorical researchers also frequently encounter such issues and have proposed concepts such as ``circulation'' and ``iconographic tracking'' from a materialist approach~\cite{gries2013iconographic}, which encourage the study of how images are produced and interpreted as they circulate.
This epistemology allows us to shift our focus from revealing the complete historical truth to understanding the octopus maps that are preserved on the Internet today.

Thus, we sought to collect a corpus of octopus maps from the web following the guidelines of iconographic tracking~\cite{gries2013iconographic}, which contains four typical phases (see \cref{fig:criteriaOct}).
\textbf{(i) ``Data hoarding''}, which means collecting as much data as possible using a macro-scaled, digital approach, such as using basic search engines to identify relevant images. In our case, we searched for keywords such as ``octopus map'' on Google Images and several well-known map databases (\ie David Rumsey Map Collection~\cite{cartography_association_david_nodate}, PJ Mode Collection~\cite{mode_persuasive_2025}, Library of Congress' Collection with Map~\cite{library_of_congress_collections_2025}, and University of Illinois Urbana-Champagne Map Library~\cite{university_of_illinois_urbana-_champaign_library_map_nodate}). At this stage, we collected 24 octopus maps, most of which are classic designs (\eg \cref{fig:criteriaOct} a).
\textbf{(ii) ``Data mining''}, which involves sorting through the collected data to identify patterns, trends, and relationships. A technique that can be employed in this phase is ``generating key terms, or tags'' to enable a ``meso-scaled'' searching. For example, after the initial search, we found that some octopus maps were also associated with keywords such as ``persuasive maps'', ``persuasive cartography'', ``propaganda'', and ``war map''. Meanwhile, we also discovered that some scattered octopus maps exist outside of large map databases, such as on map enthusiasts' communities. Therefore, we marked down this information to help refine the search for the next round. At this stage, the dataset expanded to include 53 images, featuring more diverse content and designs (\eg \cref{fig:criteriaOct} b).
\textbf{(iii) ``Assembling a collection''}. In this phase, we conducted further searches using the refined keywords and extended our search to additional sources, including Wikimedia Commons~\cite{wikicommons_octopus_propaganda_nodate}, Reddit's MapPorn community~\cite{mapporn_nodate}, and blogs~\cite{neverwasmagOctopusPolitical,teaching_with_tentacles_2013,barronmapsBarronMaps_2015,utexasVizVisual}, including more contemporary samples (\eg \cref{fig:criteriaOct} c). With these refined keywords, we accumulated a dataset of 132 images. Up to this phase, our focus had been on including as many samples that matched the keywords as possible. Our inclusion criteria specifically targeted samples that aligned with our keywords. While we did not set an explicit exclusion criterion, we excluded results that were clearly irrelevant or nonsensical.
\textbf{(iv) ``Data Cleaning.''} During this phase, a set of formal exclusion criteria were established. 
For example, a significant noise we noticed was the presence of purely political cartoons, where designers used octopuses to satirize evil political figures or countries without encoding any data. Therefore, we cleaned the corpus according to: (a) The image should encode geographic data, such as locations, borders, and topological information. Thus, images that do not contain any geographic information were excluded (\eg \cref{fig:criteriaOct} d). 
(b) The image should contain one or more octopuses. Thus, images that focus on other animal metaphors (\eg spiders) were excluded. Notably, we encountered a few corner cases which do not depict a very clear octopus image but possess many octopus-like characteristics, such as out-reaching and twisted lines. For these cases, we did not directly exclude them, because, as indicated by Puerta~\etal~\cite{puerta2025many}, an octopus map can be ``implicit'' sometimes. However, if we were to consider any flow map with radiating branches as an octopus map, that would be overly inclusive. Therefore, we adopted a compromise approach: if the description text of these maps has been explicitly mentioned, the design implies an octopus metaphor, and we included them in our corpus (\eg \cref{fig:criteriaOct} b); otherwise, we did not. 

As a result, we identified a total of \ncorpus qualified octopus maps. To double-check, we also conducted a reverse verification to ensure that our corpus covers diverse and relatively comprehensive samples. Specifically, we compared the images in our corpus with the cases mentioned in prior studies of octopus maps~\cite{puerta2025many,tyner_persuasive_1982,mode_not_2017,brooke2018phantom,baynton-williams_curious_2015,stangl_geographic_2016,roderick_mistaken_2016,mode_not_2017,cohen_modern_2009,zanin_octopus_2021,kazamias_visual_2022}. For example, Puerta et al.~\cite{puerta2025many} collected over a dozen octopus maps through personal daily collection efforts from blog posts and university libraries. Additionally, the book \textit{The Phantom Atlas}~\cite{brooke2018phantom} devotes a chapter to discussing several exemplary octopus maps in history. We compared the images we collected with those mentioned in such literature and found that all of these images have been included in our corpus. The corpus, as well as our codes and design space, can be browsed at \url{https://octopusmap.github.io}.

\subsection{Corpus Overview}
\label{ssec:corpus_overview}

\begin{figure} [h]
    \centering
    \includegraphics[width=1\linewidth]{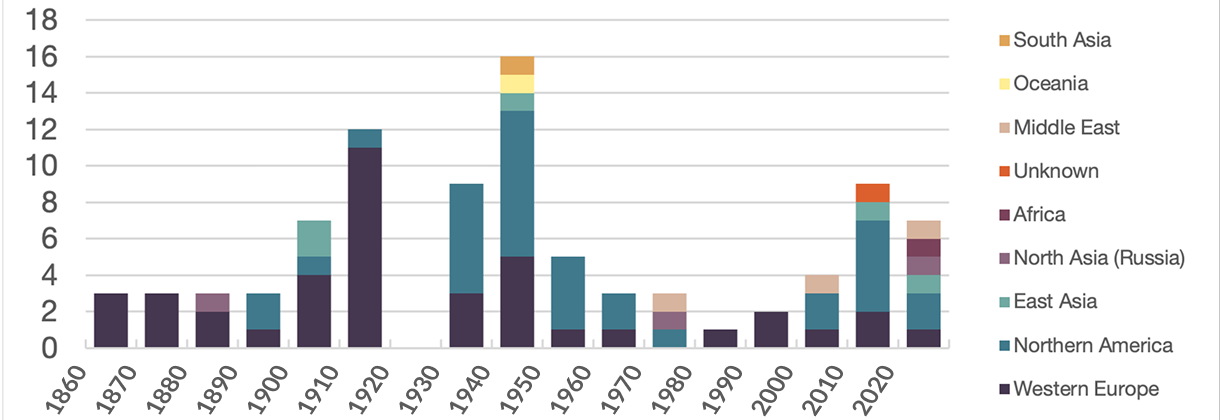}
    \caption{Distribution of publication regions over time.}
    \label{fig:Geo_Freq}
    \vspace{-1em}
\end{figure}

The final corpus contains 90 octopus maps published from 1866 to 2025. 
The earliest map in the corpus is from Victor Hugo's illustration for his novel \textit{Toilers of the Sea}, while the most recent one was collected from Instagram.
\cref{fig:Geo_Freq} shows the distribution of these maps along a timeline divided into intervals of 10 years each. Before 1900, the number of octopus maps is relatively small. However, starting around the 1910s (World War I), the number of octopus maps surged. From 1940 to 1950 (World War II), another peak emerged.
In the earlier stage from 1860 to 1920, the main publication area for these maps was Western Europe. Since the 1940s, the primary place of publication shifted to North America. The diverse distribution of octopus maps and their fluctuations in response to international events and political power dynamics provide vivid evidence of image circulation.


Based on the corpus, we would also like to highlight the following observations that may help clarify the actual situation and latest trends of octopus maps:

\textbf{Octopus maps, although often reported as a historical phenomenon, are still being produced, mainly through the internet.} We found 20 samples published after 2000, with most coming from online media outlets (9) and social media bloggers (8).
For example, \textit{Trump's Tie} was published on the author Kenneth Field's blog directly, with an blog post introducing the map-making process~\cite{Field_Trump_2017}.


\textbf{The theme of octopus maps has always been adapted to serve the latest persuasive agenda.} The  specific content in octopus maps is constantly updated to fit latest trends and events. While many early octopus maps carried the propaganda tasks of World War I and World War II, more recently, contemporary political agendas have been injected into such designs. For instance, we found several maps that metaphorically depicted US President Donald Trump as an octopus, as well as several octopus maps satirizing Russian Prime Minister Vladimir Putin.

\textbf{In recent years, the creators of octopus maps have become increasingly diverse.} As shown in \cref{fig:Geo_Freq}, although Europe and North America have been the main sources of octopus maps throughout the twentieth century, in the past two decades, we have also found a number of works from non-Western regions such as the Middle East, East Asia, and Africa. For example, the map \textit{Octoputin} satirizes Russian ambitions in Ukraine~\cite{Day_Octoputin_2014}, and the map \textit{Secure borders for Israel} presents the theme of the Palestinian-Israeli conflict~\cite{kazamias_visual_2022}.

This corpus of octopus maps serves as a foundation for our subsequent design analysis in \cref{sec:framework}. We also encourage researchers to utilize this corpus to perform further studies in the future.


\section{Analysis Framework and Corpus Coding}
\label{sec:framework}

For a long time, the field of rhetoric has accumulated a considerable amount of theoretical work in studying various communicative media, forming a specialized subfield known as visual rhetoric, which primarily studies visual media such as pictures, photographs, and videos. The octopus maps that this work investigates are also a type of visual media, and their rhetorical analysis can be effectively guided by existing interdisciplinary resources. To this end, in this section, we will first introduce the theoretical framework we constructed and adapted for analyzing octopus maps, based on literature about visual rhetoric. We will then describe how we coded our corpus guided by this framework.


\subsection{Rhetorical Schema Theory}
\label{ssec:framework_theory}

The study of visual rhetoric in images has different entry paths. One thread of rhetorical analysis focuses on understanding the author's intentions, that is, the purposes behind the creation of visuals and how these purposes relate to the historical context, social environment, and personal experiences. In contrast, another thread does not aim to trace the author's intentionality but instead pays more attention to the functions of visuals. A leading figure in this area is Foss, who proposed the rhetorical schema theory~\cite{foss_theory_2004}.
This theory advocates analyzing how an image achieves meaning-making and persuasion through the manipulation of various visual elements.
Over the decades, it has been widely applied in the analysis of various visual artifacts, particularly those related to political topics, such as propaganda maps~\cite{pickles_texts_1992} and the ``visual events''~\cite{brunner_argumentative_2016} on social media. These characteristics make the theory well-suited to the research goals of this work.


\subsection{Coding Framework}
\label{ssec:framework_dimensions}

Guided by the theory, we developed a coding framework to analyze how visual elements are manipulated to serve persuasion.

\subsubsection{Conceptual Metaphor} 
Conceptual metaphor examines how we structure our understanding of abstract concepts by mapping them onto more concrete, familiar domains, and is an important concept in visual rhetoric analysis~\cite{lakoff_contemporary_1993,lakoff_women_2008,pokojna_language_2025}. Given that octopus maps are rich in vivid metaphors, we set conceptual metaphor as the first coded dimension and analyzed three common types of conceptual metaphors:

\textbf{Ontological metaphor} refers to an abstract target domain using a source domain as a physical entity, allowing people to quantify and identify a more abstract concept. For example, when we say, ``his ego is fragile'', we use the metaphor ``ego is an entity'', where the ego is the target domain and the entity is the source domain. In octopus maps, the octopus obviously serves as an ontological metaphor.

\textbf{Structural metaphor} refers to a target domain using a source domain with a similar conceptual structure. For example, in the metaphor ``argument is war'', ``argument'' is the target domain we are less familiar with, while ``war'' is the source domain that is more familiar to us. This metaphor highlights the similar structure between war and argument. This dimension is crucial because octopus maps not only depict octopuses but also portray their behavior, demeanor, and interactions, potentially containing rich structural metaphors.

\textbf{Orientational metaphor} involves the spatial perception of human beings, locating a target domain with spatial terms such as up and down, central and peripheral, deep and shallow. In visual images, orientational metaphors are common. For example, ``left is the past'' and ``right is the future'' are common orientational metaphors~\cite{hullman_visualization_2011,kress_reading_2020}. In cartography, for example, the choice of orientation, such as what to place in the center~\cite{barton_ideology_1993,welhausen_power_2015} and what to put on the periphery~\cite{barton_ideology_1993} (in other words, where the map ends), is found to be rhetorical.


\subsubsection{Rhetorical Strategies} 

While conceptual metaphor underpins \textit{what someone wants to convey} through an image, we also analyzed \textit{how they convey it}, which involves a range of more specific rhetorical strategies and techniques. Given the complex elements involved in octopus maps, we decomposed an octopus map into three editorial layers~\cite{hullman_visualization_2011,shi2021communicating}: cartographic elements, octopus image, and text. Within each layer, we consulted existing literature to identify potential coding dimensions.

\textbf{Octopus image.} The octopus, as the most prominent imagery on the map and the object/protagonist that distinguishes it from other ordinary maps, how its image is rhetorically depicted and utilized is an important question. Visual rhetoric has a profound accumulation in understanding such figurative elements, and the coding taxonomy of political cartoons~\cite{medhurst_political_1981,morris_visual_1993} has been adapted to code octopus imagery. This is because octopus maps are a typical form of political propaganda visuals, and both political cartoons and octopus maps use vivid characters to construct a persuasion. Specifically, we divided the rhetorical analysis of octopus imagery into dimensions including \textit{contrast}, \textit{enargeia}, \textit{personification}, and \textit{placement}. 
For example, contrast denotes the deliberate juxtaposition of opposing elements to highlight differences and create emphasis, a technique common in both linguistic and visual rhetoric.
Enargeia is a rhetorical term that refers to how something is vividly recreated, corresponding to the detail level in which the octopus is depicted~\cite{kostelnick_re-emergence_2016}. Personification~\cite{medhurst_political_1981,kostelnick_pervasive_2019}, on the other hand, refers to the emotions attributed to the octopus, such as anger or envy. Placement focuses on whether and how the octopus is fused with the map.


\textbf{Cartographic elements.} Geographic information is the data foundation of octopus maps and also an important feature that distinguishes octopus maps from ordinary political cartoons. For this reason, in the second coding layer, we mainly focus on the design of the map. The codes in this part mainly refer to previous studies of persuasive cartography and critical cartography~\cite{barton_ideology_1993,monmonier_how_2018,tyner_persuasive_1982,harley2008maps}. By synthesizing elements from previous literature, we coded dimensions including \textit{map choices}, \textit{distortion by projection}, \textit{scientific symbols}, and \textit{color choices}. For example, projection is often identified as a visual rhetorical technique that helps realize distortion~\cite{monmonier_how_2018}, such as exaggerating specific areas. 
We also paid attention to the scientific symbols on the map, such as grid and legend, as these elements have been found to contribute to rendering a sense of scientific and objective feel to an image and are also commonly used techniques for persuasion~\cite{barton_ideology_1993,welhausen_power_2015}. 



\textbf{Text.} In addition to visual elements, text can also play a powerful persuasive role in maps. As Pokojná ~\etal~\cite{pokojna_language_2025} indicated, text can cooperate with visual metaphors and ``pinpoint the context.'' Pickle~\cite{pickles_texts_1992} investigated the symbol systems in propaganda maps and argued that ``maps have the character of being textual in that they have words associated with them.'' Therefore, in visual rhetoric research, the relationship between text and image is also a classic and important issue to be examined. Drawing upon relevant literature~\cite{pickles_texts_1992,tyner_persuasive_1982}, we coded two main dimensions of the text, including how it achieves \textit{explanation} and its \textit{language choice}.



To sum up, by consulting existing literature on visual rhetoric, we constructed a coding framework consisting of two dimensions: conceptual metaphors~\cite{lakoff_metaphors_2008,lakoff_contemporary_1993} and rhetorical strategies~\cite{hullman_visualization_2011,monmonier_how_2018,barton_ideology_1993,tyner_persuasive_1982,pandey_persuasive_2014}. And within each dimension, we derived a set of sub-dimensions to be coded. All the dimensions in this framework have their grounding in academic literature on visual rhetoric.
To ensure the reliability and appropriateness of the coding framework, the authors engaged in four rounds of discussion, debate, and iteration. We also consulted two professors specializing in visual rhetoric research to refine our framework and wording.
After this, we adjusted some of the codes. For example, \textit{contrast} was originally coded under the category of \textit{octopus image}. However, the experts suggested that contrast could exist throughout the entire image, meaning it is a universal rhetorical strategy. Therefore, we extracted codes that could be applied globally and created a new category called \textit{universal strategies}. The specific sub-dimensions include \textit{contrast}, \textit{metonymy}, and \textit{display}.




\subsection{Coding Process}
\label{ssec:framework_coding}


Three researchers coded the 90 samples in our corpus. One researcher is an expert in visual rhetoric, while the other two had undergone prior training on the dimensions and concepts, and thoroughly understood the key dimensions and concepts in our coding framework. 
In general, we adopted a deductive coding process based on the initial framework~\cite{peterson_rhetorical_2001}, performing close readings of the images and generating codes. To ensure our framework did not overlook important dimensions and to refine the codes when necessary, we also wrote memos to record emergent insights as a complementary deductive procedure~\cite{creswell_qualitative_2016}. Each coder coded independently without being influenced by the other coders. At the early stage, we coded the images from the data hoarding stage and found that the agreement rate between coders was not satisfactory, with a Krippendorff's $\alpha$ = 0.67.
Therefore, the coders met to discuss and resolve disagreements,
as well as to share their memos and insights, and to propose suggestions for refining the codes. For example, while most codes were found to be relatively objective and unambiguous, some were more complex, prompting us to establish more concrete criteria for these codes during our discussions. For instance, to assess the enargeia of an octopus, we referred to Boy et al.'s work~\cite{boy2017showing} and categorized it into three levels: realistic, iconic, and abstract. Besides, when coding colors, we found that different coders had significant differences in their interpretation of colors. For example, for a map that is black, one coder thought black is a common and objective color, while another coder felt that black seemed to imply the downfall of the land. In our discussions, we reached a consensus that we, as coders, should not over-interpret colors but only code those that are clearly semantically meaningful. For example, if an image binds red to blood to create a sense of horror, then we coded this color as a meaningful design.
Another complex code was the map projection. To examine whether there was distortion in the projection, we manually checked each map. 
Specifically, we overlaid each sample with various projections referring to the categories derived from previous studies on persuasive cartography~\cite{monmonier_how_2018,barton_ideology_1993}, especially the most frequently used three standard projections (Mercator projection, Peter Gall's Projection, Winkel Tripel projection) in Adobe Photoshop to determine the projection used and to identify any distortions in area size, angle, distance, and direction, as suggested by Monmonier~\cite{monmonier_how_2018}. This approach, however, represents a necessary trade-off, as there are hundreds of map projections in existence, making it impractical to manually compare each one. While automated methods for reverse-engineering projections from images exist, they often lack adequate precision, especially for historical images with pixel loss and damage. Given these challenges, we limited our analysis to several most common projections. For maps whose projections could not be verified, we tagged them as unsure and temporarily excluded them from our distortion analysis. After several rounds of discussion addressing these disagreements, we further refined the codes and aligned the coder's understanding to reach a Krippendorff's $\alpha$ = 0.82~\cite{Krip2019Content}.

\section{Unveiling the Design of Octopus Maps}
\label{sec:space}

\begin{figure*}[t]
    \centering
    \includegraphics[width=1\linewidth]{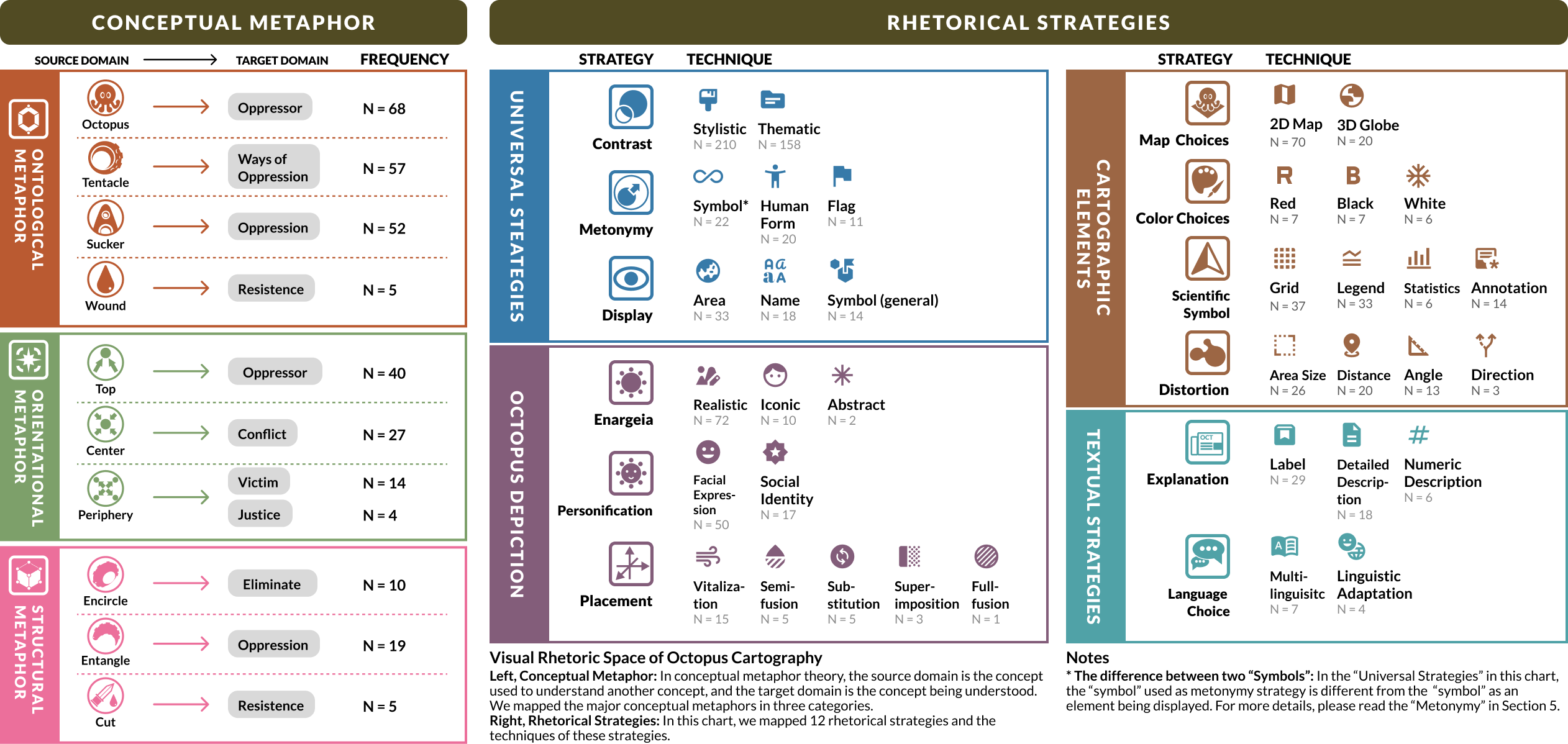}
    \caption{This figure illustrates the visual rhetoric of octopus maps. As reported in \cref{sec:space}, the analysis framework consists of two major dimensions: conceptual metaphors (illustrated on the left side) and rhetorical strategies (illustrated on the right side). Within each dimension, we present its subcategories and their frequencies. An online version of the design space can be browsed at \url{https://octopusmap.github.io}.}
    \label{fig:designSpace}
    \vspace{-1em}
\end{figure*}

Based on the coding results, we illustrate a design space of the visual rhetoric of octopus maps (\cref{fig:designSpace}). Below, we present our findings. 

\subsection{Conceptual Metaphors}
\label{sec:space_metaphor}

For octopus maps, conceptual metaphors lay the groundwork for the meanings they aim to construct and the key messages they want to express. Our analysis revealed that conceptual metaphors can be divided into three types—ontological, orientational, and structural—each containing common metaphors used to build meaning.

\begin{figure*}[t]
    \centering
    \includegraphics[width=1\linewidth]{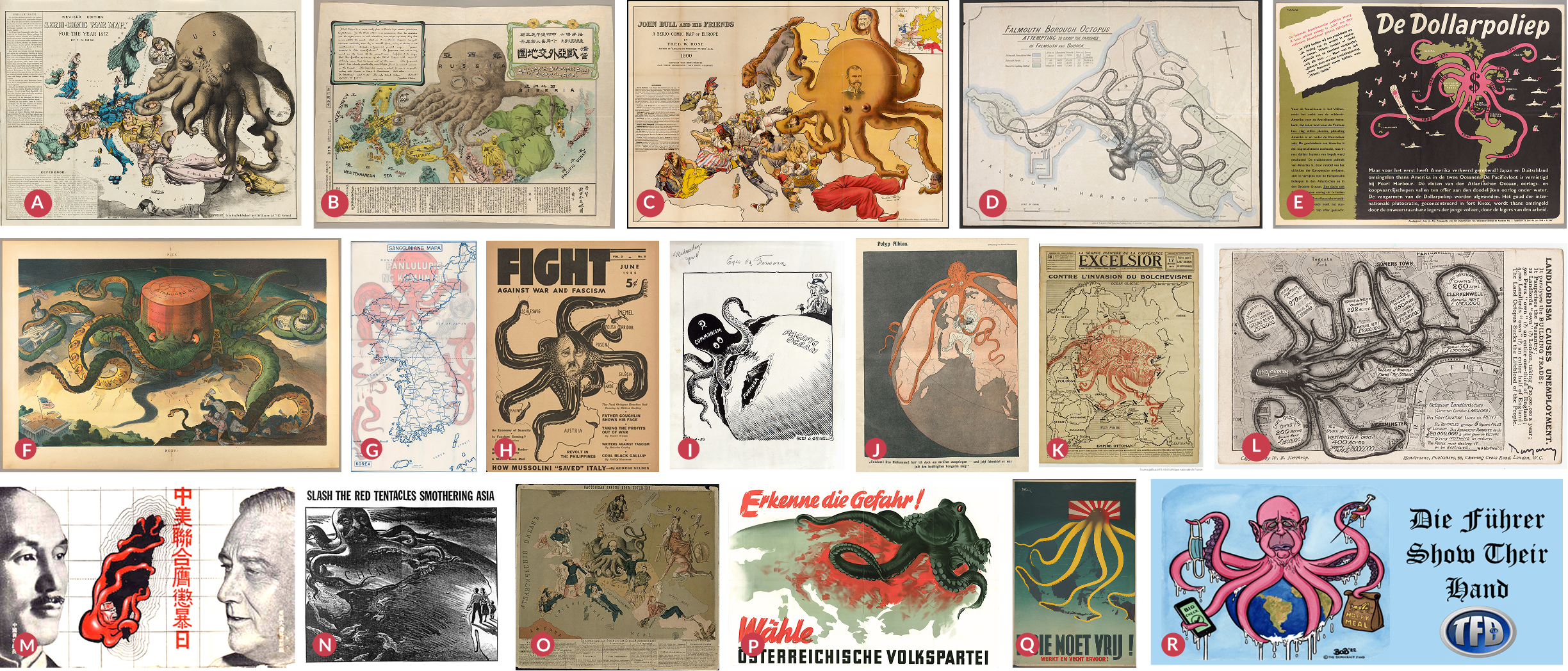}
    \caption{Examples of octopus maps in our corpus.}
    \label{fig:Plates}
    \vspace{-1em}
\end{figure*}

\textbf{Ontological metaphor.} Ontological metaphors use tangible entities and substances to refer to abstract concepts, making it the most frequently used metaphor type in octopus maps. We found that almost all octopus maps adopt the metaphor \textit{the oppressor is the octopus}. 
Upon closer examination, within the metaphorical system of \textit{the oppressor is the octopus}, the specific ``oppressor'' can vary from map to map. It could be the \textit{invader} (N = 32), the \textit{hostile ideology} (N = 22), or the \textit{financial exploiter} (N = 14). We also identified a set of associated metaphors, such as \textit{the ways of oppression are the tentacles}, \textit{oppression is the sucker}, and \textit{the successful resistance creates the wound on the tentacles}. For example, Rose's famous \textit{Serio-Comic War Map from 1877} (\cref{fig:Plates}, A) uses an octopus to depict Russia as the invader. Each tentacle represents a specific military invasion of a European country. The wound on the octopus' tentacles represents Russia's failure in Crimea, with the toponym ``Crimea'' overlapping the wound clearly.


\textbf{Structural metaphor}. A structural metaphor uses a more relatable source domain to refer to a target domain with a similar mental structure. In octopus maps, \textit{to oppress is to entangle} is the most identified structural metaphor (N = 19). We can also find the metaphors, \textit{to eliminate an area is to encircle it} (N = 10) and \textit{to cut the tentacle is to resistance}. Echoing what  Puerta et al.~\cite{puerta2025many} called ``grabbiness'', in octopus maps, the depiction of entanglement implies that ``some influence over the objects in the grasp of its tentacles.'' In most cases, the entanglement is literal and described vividly. For example, \textit{The Dollar Octopus} (\cref{fig:Plates}, E) depicts a giant American octopus plundering the colony at the time. Each tentacle entangles different regions, including Hawaii, Alaska, the Philippines, and Iceland.

\textbf{Orientational metaphor}. Orientational metaphors generate meaning through the manipulation of spatial relationships. We identified two main types of orientational metaphors in octopus maps. The first is \textit{the oppressor is at the top} (N = 40). For example, the map \textit{Recognize the danger! Choose the Austrian People's Party} (\cref{fig:Plates}, P) places the octopus (invader) in the upper right corner, conveying a sense of dominance and looming threat.
The second type is \textit{the conflict is at the center} (N = 27). For instance, the map \textit{A Humorous Diplomatic Atlas of Europe and Asia} (\cref{fig:Plates}, B) places the central conflict at the visual center, highlighting the entanglement of Poland by Russia's tentacles, thus emphasizing the core conflict.
Meanwhile, positioning elements at the periphery can also be meaningful. We found two metaphors concerning peripheral placements in octopus maps: \textit{the victim is at the periphery} (N = 14) and \textit{justice is at the periphery} (N = 4). In the map \textit{A Humorous Diplomatic Atlas of Europe and Asia} (\cref{fig:Plates} B), the central conflict between Russia and Poland is placed at the visual center of the map, representing the major conflict in Europe battlefield; meanwhile, the countries being invaded are placed at the periphery of the map, which are the places for the victims. 
In other cases, such as the maps \textit{Next!} (\cref{fig:Plates}, F) and \textit{Eyes on Formosa} (\cref{fig:Plates}, I), the White House and personified America is intentionally placed at the edge, representing America as a detached observer and symbol of justice.

\raisebox{-0.5ex}{\includegraphics[height=1em]{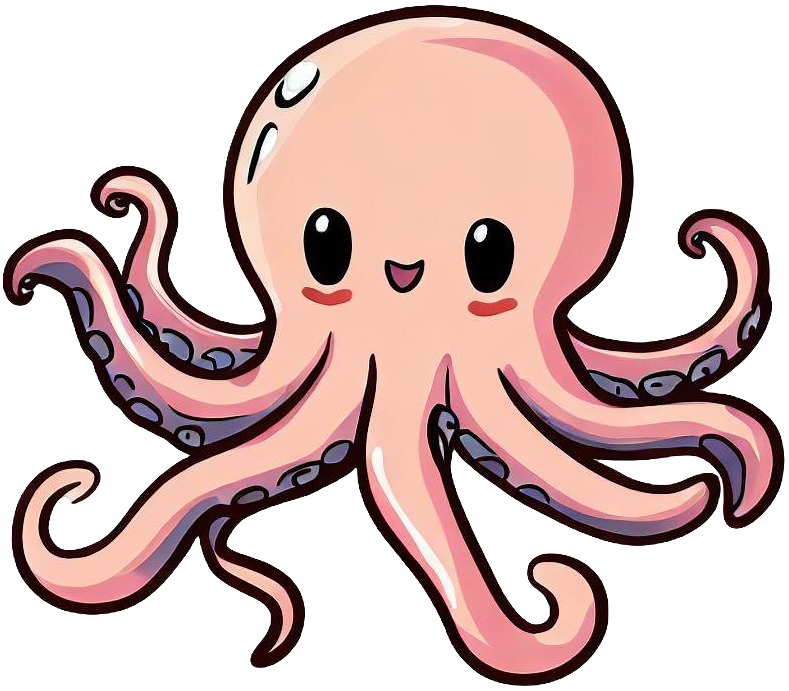}} \textbf{To sum up}, we find that a typical conceptual metaphor of octopus maps is \textbf{a metaphoric system of aggressive oppression}: The octopus represents an oppressor (ontological metaphor), typically positioned in the center or upper part of the map, looming over others in a dominant stance (orientational metaphor), while simultaneously exerting its oppressive action on others through various degrees of entanglement (structural metaphor).


\subsection{Rhetorical Strategies and Techniques}
\label{sec:space_strategies}

Under the high-level framework established by conceptual metaphors, we identified a rich set of low-level visual rhetorical strategies and techniques to further enrich and strengthen the persuasiveness of octopus maps. As introduced in \cref{sec:framework}, we categorized these strategies based on the visual elements they apply to, including universal rhetorical strategies, cartographic strategies, octopus depiction strategies, and textual strategies. Within each category of strategies, we have also identified a set of specific techniques.

\subsubsection{Universal Rhetorical Strategies}

These strategies can be applied to any element on octopus maps.

\textbf{Contrast\iffalse (N = 368) \fi:} Contrast is the most frequently used universal strategy to reinforce the visual rhetoric of the octopus maps, and one map can contain multiple contrasts. Specifically, two types of contrast can be used: \textit{thematic contrast} (N = 158) and \textit{stylistic contrast} (N = 210). The most frequently used thematic contrasts are the contrast of \textit{the invading and being invaded} (N = 22) and \textit{strong and weak} (N = 19). The most frequently used stylistic contrasts are \textit{light and dark} (N = 43), and \textit{big and small} (N = 41).
For example, we can see the contrast between the invading and being invaded in the Rose' \textit{Serio-Comic War Map}, which juxtapose the giant octopus as the invader and the other European countries as the invaded. 
The map also creates a contrast between the strength of the Russian octopus and the weakness of the countries being occupied combining with multiple stylistic contrasts, including \textit{light and dark}, \textit{big and small}.

\textbf{Display\iffalse (N = 65)\fi:} Display is an important visual rhetorical strategy about (1) “how things look or appear,” (2) exhibition or demonstration, and (3) showiness or ostentation, as Prelli~\cite{prelli_rhetorics_2021} identified. In octopus maps, certain information is displayed intentionally to get across ideological meaning. We identified three types of elements that are typically displayed to convey meaning: \textit{area} (N = 33), \textit{name} (N = 18), and \textit{symbol} (N = 14). For example, the map \textit{Nasa Pagtankilik ng Nagkakaisang Bansa} (\cref{fig:Plates}, G) displays the toponym ``Manchuria'' beside the octopus on the upper left blank space, annotating the northwestern China. The display of this toponym could entail the ideology of the map maker: in 1953, when this map was made, the toponym of Manchuria was no longer used in China as ``it was too closely associated with the Manchukuo state.''~\cite{gamsa_manchuria_2020} Displaying Manchuria as the name implies the entity of the enemy is from the northeastern China. Another example for displaying symbol and area is the octopus map, \textit{Next!} (\cref{fig:Plates}, F). This map describes how the Standard Oil Company controls the U.S. government. This map only displays the area of Washington D.C, with the buildings of the white house, the state house, and the capitol, reflecting the main focus on the financial control of the U.S. government at the time. 

\textbf{Metonymy\iffalse (N = 53)\fi:} Metonymy is the abstraction of a concept or idea. For example, \textit{death is a skull} is a metonymy, in which the skull is included in the death process. In octopus maps, the creators usually use a \textit{symbol} (N = 22), \textit{human form} (N = 20), and \textit{flag} (N = 11) to represent a country. The function of metonymy is to indicate the semantics of the octopus, allowing audiences to understand who is the enemy quickly. For example, \textit{Fight Against War and Fascism} (\cref{fig:Plates}, H) uses the face of Hitler as a metonymy that refers to Fascism. \textit{Eyes on Formosa} (\cref{fig:Plates}, I) uses the symbol of a sickle and hammer to refer to communism and its confrontation with the US during the Cold War.


\subsubsection{Cartographic strategies}

Cartographic strategies are those applied to maps.

\textbf{Map choices:} There are two categories of map choices: the \textit{3D globe} (N = 20) and the \textit{2D map} (N = 70). The choice of the map types serves as the starting point of the persuasive function and entails different view points that create different sense of objectivity~\cite{kennedy_work_2016}. Specifically, the choice of map is also associated with the choice of projection. Two-dimensional (2D) maps always distort geographical information to some extent~\cite{monmonier_how_2018}. However, while 3D maps might convey a more accurate sense of spatial information, at least in static images, they still leave ample room for rhetorical emphasis and concealment. For instance, the map \textit{Octopus Albion} (\cref{fig:Plates}, J) contains a 3D globe that creates a sense of a worldwide view. However, it only displays Europe as the center and conceals the United Kingdom in the 3D globe representation.

\textbf{Distortion:} Four types of map distortion are found in our corpus: \textit{area size} (N = 26), \textit{direction} (N = 3), \textit{distance} (N = 20), and \textit{angle} (N = 13). Some octopus maps use cartographic elements more akin to political cartoons, with a rather careless depiction of geographical scale. For example, the area size and distance are distorted on the map \textit{Slash the Red Tentacles} (\cref{fig:Plates}, N). In this design, the Pacific Ocean appears strikingly shrunken, making the octopus representing Stalin look overwhelming and exaggerating its threat to America. What perhaps is more noteworthy is that distortion can also be found in octopus maps that use more ``professional map-making techniques'', such as seemingly rigorous projections and graticules. For instance, the famous map \textit{John Bull and His Friends} (\cref{fig:Plates}, C) includes a smaller scientific-style map in the upper right corner, aiming to convey more accurate geographic information compared to the personified map. However, we still found distortion in the area size on this map: compared to the standard Winkel Tripel projection, the area size of Russia is reduced, especially in eastern Russia. In other word, it uses scientific symbols to reinforce its persuasive power rather than providing accurate geographic information.

\textbf{Color Choices:} The colors used to distinguish or annotate different areas in maps are different from those for other visual elements. As Bertin noted, suitable color use guarantees the selectivity of geographical data~\cite{bertin_matrix_2000}. In octopus maps, we found that red (N = 7), black (N = 7), and white (N = 6) are the colors most intentionally bound with certain meaning. Echoing what Monmonier noted, the red could be used to depict the former Soviet Union, Cuba, or China in a propaganda map; black can represent being occupied or death, and white can symbolize cleanliness. Likewise, Kimball analyzed Charles Booth's \textit{Maps of London Poverty} and indicated that the black color on the map ``will show us where to find the haunts of the lowest class.'' 
A typical example of intentional color choice is the visualization on an anti-communism pamphlet (see \cref{fig:criteriaOct}, b). In this map, the USSR is colored in red, other areas of the world are grey, and only the United States is colored in white. The white color functions not only to highlight but also to imply the cleanliness of the area.

\textbf{Scientific Symbols:} Using scientific symbols in octopus maps builds the \textit{ethos} (appealing to credibility) to strengthen the sense of objectivity and reliability. In our corpus, four types of scientific symbols have been applied: \textit{grid} (N = 37), \textit{legend} (N = 33), \textit{annotation} (N = 14), and \textit{statistics} (N = 6). For example, \textit{Falmouth Borough Octopus} (\cref{fig:Plates}, D), which attempts to grasp the parishes of Falmouth and Budock, provides detailed statistics and legends on the map. The statistic table consists of the acreage, population, rateable values, loans, and liability figures from the 1881 census of Great Britain. This professional data not only reveals the details of the land annexation at the time but also tries to build the evil octopus upon the touch of facts using scientific discourse.

\subsubsection{Octopus Depiction Strategies}
\label{sssec:octopus_depiction}

These strategies focus on the representation of the octopus.

\textbf{Enargeia:} Enargeia, or vivid description, is a common rhetorical strategy expected to elicit emotions and reinforce persuasion~\cite{kostelnickHumanizingVisualDesign2019}. In our corpus, the \textit{realistic} depiction of the octopus image is dominant (N = 72), while the \textit{iconic} octopus (N = 10) and the \textit{abstract} octopus (N = 2) occasionally appear. In most of the octopus depictions, the suckers are crafted in detail. 
For instance, the creator arranges the suckers according to the spatial change precisely on the map \textit{Octopus Albion} (\cref{fig:Plates}, J), rendering the evilness of the octopus. Moreover, the vivid description provides more space for meaning-making, especially for conceptual metaphors such as \textit{tentacle is the oppression}, \textit{entanglement is the action of grasping}, and \textit{cutting is defeating evil}. 

\textbf{Personification:} Personification prevails in the octopus maps. The technique of personification includes depicting \textit{facial expressions} (N = 50) and representing \textit{social identity} (N = 17). The creators depict facial expressions with rich emotions, intending to persuade audiences by appealing to emotions, and they represent social identity to tell the audiences \textit{who} owe the emotions. For instance, the creator assembles a realistic Hitler portrait on the map in the pamphlet \textit{Fact Spy Stories Pulp} (\cref{fig:Plates}, H). In this map, Hitler is depicted as a cruel politician with eyes staring at the earth. Using this technique, the information about who the threat is and what the threat looks like is delivered to the audience. Using social identity as the technique of personification can be found in using the costume to represent the social character. For example, in \textit{The Real Europe Without Make-up} (\cref{fig:Plates}, O), the Russian octopus is depicted as a commander in the costume. This personification technique further locates the octopus metaphors in a socio-cultural context. 


\textbf{Placement:} We identified five types of strategies regarding octopus placement: \textit{vitalization} (N = 15), \textit{semi-fusion} (N = 5), \textit{substitution} (N = 5), \textit{superimposition} (N = 3), and \textit{full-fusion} (N = 1). Vitalization refers to the specific strategy in which the octopus entangles the land and creates spatial overlap to vitalize the land as a physical entity. For instance, the Hitler octopus entangles the land of the Polish Corridor and separates Memel on the map \textit{Fight Against War and Fascism} (\cref{fig:Plates}, H). With a vivid representation of the spatial overlap, the creator uses this strategy to create an interaction between the cartographic data and the pictorial octopus, reinforcing the argument about how Hitler annexed the land of another country. Semi-fusion technique allows the map makers to represent the borderline of a country as a tentacle. On the map, \textit{The Indies Must Be Free! Work and Fight For It} (\cref{fig:Plates}, Q), the tentacles of the octopus fuse with the borderline of Indonesia, representing that the Japanese Empire has controlled the territory. The substitution technique replaces the borderline with the octopus. For example, the borderline of Russia is hidden while the human forms personification resembles the borderline on the \textit{Serio-Comic War Map For The Year 1877} (\cref{fig:Plates}, A). The map maker conceals and replaces the territory with a dark, giant octopus. The superimposition technique allows the octopus to overlap the octopus on the map without losing cartographic information. Interestingly, we found that many cartographers like to use such techniques to enhance the fusion between cartographic elements and the octopus. For instance, \textit{Against the Invasion of Bolshevism} (\cref{fig:Plates}, K) uses the capital symbols of St. Petersburg and Moscow to represent the eyes of the octopus, expressing the connotation that these two cities of Russia are the eyes of the octopus.

\subsubsection{Textual Strategies}

Textual strategies manipulate the text on octopus maps.

\textbf{Explanation:} The textual layer of the octopus maps often provides rich explanations for persuasion. The common techniques for explanation are \textit{label} (N = 29), \textit{detailed description} (N = 18), and \textit{numeric description} (N = 6). The label is the most common technique to explain the meaning of the octopus. The earliest example is \textit{The Real Europe Without Make-up} (\cref{fig:Plates}, O), in which we can see the map maker labels the octopus with the name of Prussia and each warfare. Moreover, map makers also use detailed descriptions to describe the villain as an octopus and provide background information about the war. For instance, the second edition of the \textit{Serio-Comic War Map For The Year 1877} (\cref{fig:Plates}, A) provides a detailed description titled ``explanation'' in the upper left corner, with a depiction of the Russian octopus: ``The northern colossus—Russia—is represented in the form of a wild-looking octopus...''  Following this, the map maker uses 7 paragraphs to introduce the situation of the other countries on the map. The use of statistics is another technique to explain the context of the octopus map. For instance, in \textit{Falmouth borough octopus attempting to grasp the parishes of Falmouth and Budock} (\cref{fig:Plates}, D), the map maker cites the data from the 1881 Census of Britain and composes a table to explain the quantitative information of each parish. As another example, to criticize the excessive control of landlords over land, \textit{Landlordism Causes Unemployment} (\cref{fig:Plates}, L) explains the size and rental price with textual annotations on the corresponding areas on the map.

\textbf{Language Choice:} The choice of language entails the consideration of the rhetorical situation, including the audience and the context. We identified the techniques of language choice below: \textit{multi-linguistic} (N = 7) and \textit{linguistic adaptation} (N = 4). Multi-linguistic is the most commonly used technique to expand the circulated areas. For instance, the second edition of \textit{The Serio-Comic War Map} (\cref{fig:Plates}, A) includes the ``new and more extensive bi-lingual key in both English and German''~\cite{roderick_mistaken_2016}. Linguistic adaptation considers the effect of propaganda and selects the language of the target country. For instance, the \textit{flyer No.107} (\cref{fig:Plates}, M), printed by the Office of War Information of America, chooses traditional Chinese as the language in order to circulate in Taiwan.

\raisebox{-0.5ex}{\includegraphics[height=1em]{Figures/icon.png}} \textbf{To sum up}, we infer that the strong appeal and long-lasting influence of octopus maps are closely related to their carefully selected and meticulously crafted visual rhetorical design. Their effectiveness stems not merely from the octopus motif but from the synergistic interplay of all visual elements—cartographic features, symbolic imagery, and text—each deliberately employing rhetorical devices to co-amplify persuasion. This holistic design transcends simplistic interpretations of the genre as ``maps with an octopus.''
Moreover, technically speaking, we found that these maps employ highly rich and carefully planned design techniques, such as subtle cartographic distortions and seamless integration of the octopus with geographic boundaries or gridlines. Although these nuanced techniques have rarely been deeply discussed in previous research on persuasive visualization, we found that they are critical to the maps' success: they marry artistic vividness with scientific rigor, enabling the design to convey data-driven arguments with both emotional resonance and analytical credibility.
Ultimately, octopus maps exemplify how layered visual rhetoric and calculated design choices forge compelling, enduring persuasive artifacts.

\section{Dynamics of Visual Rhetoric}
\label{sec:dynamics}

Although we have outlined a rhetorical design space in the previous section, we don't claim that the identified methods and rhetorics are universally applicable. In fact, during the process of collecting and analyzing the corpus, we discovered that the interpretations of rhetoric can vary across different cultures and contexts. 
For example, we calculated the co-occurrence of the source regions of the octopus maps and the conceptual metaphors they used. As the results suggest, most maps are from Western Europe (N = 41), and most octopuses in these images refer to ``invaders'' (N = 21) and ``financial oligarchy'' (N = 8). Images from North America (N = 34) frequently describe octopuses as ``invaders'' (N = 13) as well, while they are more concerned with ``communism'' (N = 11) during the Cold War.
However, while most octopus maps convey negative political meanings, exceptions do exist: several octopus maps depict octopuses with positive or unconventional connotations. We present case analyses below.

\begin{figure}[h]
    \centering
    \includegraphics[width=1\linewidth]{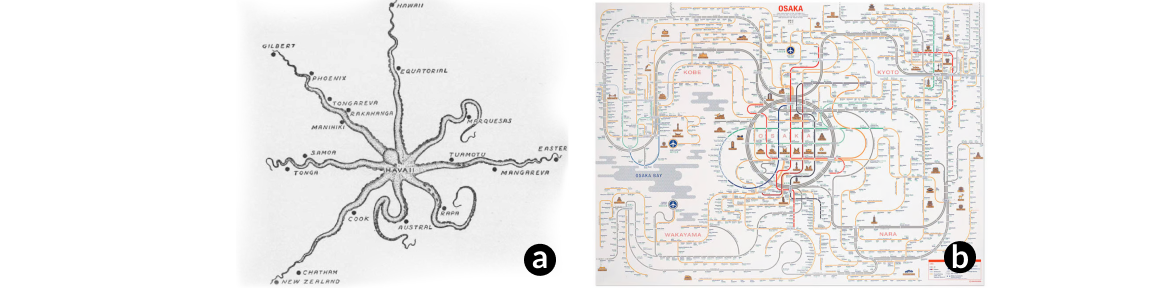}
    \caption{(a) Vikings of the Sunrise, (b) City map: Osaka. }
    \label{fig: flux_case}
    \vspace{-2em}
\end{figure}

\subsection{Case I: The Polynesian Octopus Map}
\label{ssec:dynamics_case1}

As shown in \cref{fig: flux_case} (a), this is an octopus map called \textit{Vikings of the Sunrise} archived at the Kaiwakiloumoku Hawaiian Cultural Center~\cite{Hawaii_Cultural_Center_2019}. It was made by Dr. Peter H. Buck (Te Rangi Hiroa) in 1938 and showcases the Polynesian universe with Havai'i (Ra'iātea) as the center, representing the octopus metaphor of Ra'iātea, a highly sacred center in Polynesian culture.

This design shares many rhetorical strategies as identified in our design space. For example, we can identify the use of techniques such as \textit{display} (\eg the selection of the important and sacred location in Polynesian universe), \textit{distortion} (\eg inaccurate physical distance and direction), \textit{placement} (\eg the boarder of Havai'i is substituted by the octopus), and \textit{enargeia} (detailed texture, tentacles, and the suckers).
If we explain this octopus map without considering the cultural context, we might interpret Hawaii as an evil villain. However, the meaning of this metaphors is the opposite. 
The key difference lies at the level of conceptual metaphors: combining \textit{the sacred center is an octopus} (as the ontological metaphor), \textit{the sacred region is at the center} (as the orientational metaphor), and \textit{the cultural alliance is the stretch of the tentacles} (as the structural metaphor), this map in fact has an overall positive meaning.
As Dr. Buck~\cite{buck_viking_1964} says, the ``great octopus'' is frequently mentioned by the local people, which entails ``the difficulties encountered on their voyage'' and ``imbued with magic power'' in the indigenous Polynesian community. With that said, the octopus metaphor don't have negative connotation as the Western culture does but carries a religious meaning. 


\subsection{Case II: The Japanese Octopus Map}
\label{ssec:dynamics_case2}

As shown in \cref{fig: flux_case} (b), this map illustrates the metropolitan system of Osaka City in Japan. It was designed to attract more tourists and can be used as a pocket map. As introduced by its cover, this map uses the impicit metaphor of octopus: ``Osaka is closely tied to the surrounding cities of Kyoto, Kobe, Nara and Wakayama. Many people traveling to Osaka also make a stop at the neighboring cities. This map extended this concept into octopus, the Osaka metropolitan area is visualized as an octopus with the head being Osaka and the legs sprawling out to the other four cities.''~\cite{zero_octopus_2016}

Obviously, this map also share a lot of rhetorical features we discussed before (\eg \textit{display of symbol} such as the airplane, \textit{distortion} of distance to align it better with the railways, using \textit{full-fusion} to merge the octopus into the representation of the traffic system, and using \textit{detailed description} to the meaning of the octopus), however, it is grounded in different conceptual metaphors compared to those in the West. 
In the description of this case in the David Rumsey Map Collection, it depicts the octopus as ``the main ingredient of the octopus dish Osaka is known for''~\cite{zero_octopus_2016}, so that local residents' feelings towards it may not primarily be fear.
And in another Japanese map called \textit{Octopus Treading} in our corpus, the octopus is even viewed as the metaphor for victory. Therefore, as suggested by these cases, the connotation of octopus tends to be more positive in Japanese culture. 

\raisebox{-0.5ex}{\includegraphics[height=1em]{Figures/icon.png}} \textbf{To sum up}, the visual rhetoric of octopus maps can have different connotations in various cultural contexts.
Thanks to the vivid cases provided by octopus maps, we can see how different cultures interpret the meaning of visualizations even with similar rhetorical strategies.

\section{Discussion}


\subsection{Major Implications}

We summarize the major implications below:

\textbf{Octopus maps are dedicated rhetorical systems}. 
In this work, the concept of visual rhetoric has provided us with an important research perspective and guided our understanding of how octopus maps use multiple conceptual metaphors and rhetorical strategies to create a strong persuasive effect. Based on this, we propose the following guidelines for future research that also wishes to use a rhetorical lens: (i) Since the essence of rhetoric is persuasion, any visualization involving ``persuaders'' or ``means of persuasion'' can be analyzed rhetorically, such as the narrative strategies in data stories~\cite{shi2021communicating} and nudging users' attitudes using visualizations~\cite{errey2024nudging}.
(ii) Certain domains and industries, such as news media and political parties, often aim to persuade. Research related to these industries is recommended to consider a rhetorical perspective.
(iii) As said by Aristotle, rhetoric is the art of ethos, pathos, and logos. Accordingly, for example, studies on trust in visualization~\cite{padilla2022multiple} (ethos) and affective visualizations~\cite{lan2023affective} (pathos) can also benefit from a rhetorical perspective.




\textbf{Embellishment can be powerful and sophisticated.} 
According to our analysis in \cref{sssec:octopus_depiction}, we find that figurative elements can not only be a cherry on top of a cake, but also play a significant role in conveying core information and shaping the emotional atmosphere in visualizations. From a design perspective, the integration of embellishment with data visualizations can be highly diverse, ranging from simple overlay and juxtaposition to semi-fusion and deep fusion. Moreover, integration can occur not only in terms of appearance and shape but also semantically. This suggests that future user studies about embellishment should consider a wider variety of design forms and especially pay attention to more sophisticated forms such as visual metaphors.

\textbf{Visual rhetoric can be culturally dynamic.}
As discussed in \cref{sec:dynamics}, we suggest that design conventions and their persuasiveness can vary across different cultures, highlighting the dynamic nature of persuasive visualizations. 
Specifically, we examined two cases from different cultural settings. These cases demonstrate how even seemingly consistent design conventions can be interpreted differently.

\subsection{The Ethical Concerns of Persuasive Visualization}

Octopus maps not only provide vivid examples demonstrating how powerful visualization can be when used for social affairs, but also reveal the dark side of persuasive visualization through their real lessons. For instance, Fred Rose's \textit{Serio-Comic War Map For The Year 1877}, by portraying Russia as a menacing octopus reaching out to many countries and constructing a rhetoric about expansionism and threat. 
However, this design is undoubtedly political and ideological, catering to the widespread Russophobia and corresponding propaganda in England at the time~\cite{gleason1950genesis}. Ross ``crystallized, formalized, and perfected'' this political map genre~\cite{barron2008bringing}, creating a visual propaganda technique that was widely imitated and disseminated~\cite{barronmapsBarronMaps_2015}.
Today, this visual genre is still circulating on various platforms and are being rewritten with new forms of hatred and power. For example, a recent case in our corpus, called \textit{Schwabtopus} (\cref{fig:Plates} R),  depicts the vaccination policy as a pink octopus positioned at the top of the Earth to satirizes the policy during the COVID-19 pandemic. In addition, the octopus employs a strong \textit{personalization} rhetoric by drawing the face of Klaus Schwab, the founder and executive chairman of the World Economic Forum, on its head.
As it spread, widespread concern and backlash quickly arose. For example, there were requests to remove the picture due to its promotion of a ``conspiracy narrative'' and a ``hateful trope''~\cite{Diversity_2022}.
By the time of writing this paper, the image had already been removed.

Finally, we would like to emphasize that the primary goal of this research is not to assist designers in creating or reproducing propaganda maps. In particular, we do not advocate that readers of this work employ such design techniques for malicious purposes or to spread hatred and misinformation. 
Instead, we hope that our research, as well as the associated corpus and website materials, can be browsed and learned by the public, helping to enhance their visualization literacy and increase their ability to identify the potential manipulative rhetoric of visualizations. Our summarization of the rhetorical techniques may also serve as a checklist, helping data journalists and designers become aware of potential emotional overstatement and information distortion in their work. It can also inform visualization educators in developing pedagogy on how to design faithful and trustworthy data visualizations.

\subsection{Limitations}

Although we have tried our best to include as many rich samples as possible, guided by the iconographic tracking methodology, our corpus is not exhaustive. Many octopus maps may have disappeared in history, and we cannot access octopus maps in a wide range of languages. Besides, our analysis of octopus maps has primarily focused on the information presented within the maps themselves, without further considering the media context in which these maps were published (\eg the surrounding articles that appeared before and after the maps).

Another limitation of this work is that it only examines octopus maps as a specific genre of persuasive visualization. Therefore, some identified rhetorical techniques may not be fully generalizable to other types of visualizations. However, we believe that some aspects of our research may still be applicable to other relevant work. Firstly, although we focus on octopus maps specifically, the rhetorical schema theory and its associated concepts (\eg conceptual metaphor, ontological metaphor, oriental metaphor, structural metaphor) are universal and can support the analysis of various visual materials (\eg using conceptual metaphors to help understand communicative visualization~\cite{parsons2018conceptual} or to analyze infographics for scientific storytelling~\cite{pokojna_language_2025}). If researchers want to study visual rhetoric in other types of visualizations, they can also adopt this theory and make appropriate adaptations according to their research subjects.
Secondly, the coding dimensions we established for octopus maps may be reused or extended by future research that also includes cartographic elements or figurative elements. For example, the rhetoric of maps (\eg map choices, distortion by projection) is possible to appear in geographic storytelling~\cite{roth2021cartographic,li2023geocamera}. The rhetoric of octopus (\eg enargeia, personification) is likely to be observed in other embellished and metaphorical visualizations, especially those that use anthropomorphism (\eg  anthropographics~\cite{morais2020showing}).
Lastly, the iconographic tracking method, as summarized in \cref{fig:criteriaOct}, may also be applied to future studies that need to build their own corpus, especially those based on historical images.

\section{Conclusion}

Our studies provide insights into understanding the persuasive space of a famous trope: the octopus map. We collected a corpus of \ncorpus octopus maps from various online sources and. Then, drawing upon the rhetorical schema, we investigated the visual rhetoric of these maps in detail and summarized a design space. Our major contribution is a deeper understanding of the conceptual metaphors and rhetorical strategies as persuasive techniques in visualizations. Besides, through the above analysis, we also uncover a set of interesting findings (\eg the continuous impact of octopus maps on social media, the dynamics of visual rhetoric). Lastly, drawing from the lessons provided by octopus maps, we discuss the associated ethical concerns of persuasive visualization.

\acknowledgments{
This work was supported by NSFC 62402121 and Shanghai Chenguang Program. We thank all the reviewers for their valuable feedback.}

\bibliographystyle{abbrv-doi-hyperref}
\bibliography{reference}

\end{document}